\documentclass[runningheads]{llncs}
\usepackage[T1]{fontenc}
\usepackage{graphicx}
\usepackage{amsmath}
\usepackage{amssymb}
\usepackage{multirow}
\usepackage{array}
\usepackage{hyperref}
\usepackage{placeins}
\begin{document}
\title{Beyond Discrimination: Evaluating Probability Calibration in Radiomics--Image Fusion for Renal Cell Carcinoma Classification}

\titlerunning{Probability Calibration in Radiomics--Image Fusion}
%
%
%

\author{ Yuan Liang\inst{1,2}$^{*}$ \and Morgan May\inst{1,3}$^{*}$ \and Abraham Campbell\inst{2} \and Sourav Bhattacharjee\inst{4} \and Simon Caton\inst{2} \and Pierpaolo Dondio\inst{5} } \institute{ Research Ireland Centre for Research Training in Machine Learning, ML-Labs, Dublin, Ireland \and School of Computer Science, University College Dublin, Dublin, Ireland \and School of Computer Science, Technological University Dublin, Dublin, Ireland \and School of Veterinary Medicine, University College Dublin, Dublin, Ireland \and School of Computing, Dublin City University, Dublin, Ireland \\ \email{yuan.liang@ucdconnect.ie}\\ 
}

\maketitle              
\begingroup \renewcommand\thefootnote{\fnsymbol{footnote}} \footnotetext[1]{These authors contributed equally to this work.} \endgroup
\vspace{-0.2in} 
\begin{abstract}
Reliable probability estimates are important for the clinical interpretation of machine-learning classifiers, yet radiomics--image fusion models for renal cell carcinoma (RCC) are commonly evaluated primarily by discrimination. Using 396 patients from the KiTS23 dataset, we evaluated whether adding radiomics to four image-model families improved probabilistic performance in addition to discrimination. Seventeen individual models and four probability-average ensembles were assessed using discrimination, probability calibration, and proper scoring measures. Across 12 matched image-only versus fusion comparisons, AUC and AP improved in all 12, while Brier score and negative log likelihood (NLL) improved in 11 of 12 and expected calibration error (ECE) in eight of 12. The 3D MedVAE gated-fusion model achieved the strongest overall individual-model profile. Probability averaging across concatenation, cross-attention, and gated fusion produced favourable point estimates across all five primary metrics relative to image-only models in all four families, although paired-bootstrap support was stronger for Brier score and NLL than for ECE. These findings show that radiomics--image fusion can improve probabilistic prediction quality beyond discrimination, while calibration behaviour varies across image architectures and fusion strategies. We show that our proposed probability-average ensemble models improve discrimination and calibration simultaneously across all architectures in the study.


\end{abstract}
\section{Introduction}

Renal cell carcinoma (RCC) is a heterogeneous malignancy comprising several histological subtypes with distinct biological behaviour, prognosis, and treatment implications. Among RCC, clear cell RCC (ccRCC) is the most common subtype, and its preoperative distinction from non-clear cell RCC is clinically relevant for patient stratification and management \cite{bex2025european}. Contrast-enhanced computed tomography (CT) plays a central role in the assessment of renal masses, but substantial overlap in tumour appearance can make subtype characterisation challenging using visual interpretation alone. Quantitative imaging approaches have therefore increasingly been investigated to extract additional information from routine CT.

Radiomics provides predefined measurements of tumour shape, intensity, and texture and has shown promise for distinguishing ccRCC from non-clear cell RCC \cite{wang2021radiomics}. In parallel, deep neural networks can learn higher-dimensional image representations directly from CT data, while pretrained medical imaging models offer a means of exploiting such representations when labelled datasets are limited \cite{uhm2021deep,varma2026medvae}. These two feature extractions are conceptually complementary: radiomics provides structured descriptors of tumour phenotype, whereas learned image representations can capture complex spatial patterns that are difficult to specify manually. 
Consequently, integrating radiomics with learned image representations offers a potential means of improving RCC classification beyond either representation alone.

Most evaluations of such models, however, focus primarily on discrimination \cite{bhandari2021ct}. Here, discrimination refers to a model's ability to distinguish or rank ccRCC cases relative to non-ccRCC cases. Metrics such as the area under the receiver operating characteristic curve (AUC) and average precision (AP) quantify this ability, but they do not determine whether the predicted probabilities are numerically reliable. A classifier may achieve strong discrimination while systematically assigning probabilities that are too extreme or otherwise inconsistent with observed outcome frequencies \cite{van2019calibration}. This distinction is particularly important in medical AI, where model outputs may be interpreted as estimates of disease probability rather than merely as ranking scores.

Probability calibration therefore provides a complementary dimension of model evaluation. A well-calibrated classifier should, for example, assign an event probability close to 0.8 to groups of patients in whom the event occurs approximately 80\% of the time. Neural networks may exhibit substantial miscalibration despite high predictive accuracy \cite{guo2017calibration}. 
Although probability calibration has been considered in multimodal imaging studies for other clinical tasks~\cite{nelson2026gaussian,xi2020deep,raposo2026imaging,hu2026robust}, systematic evaluation of calibration remains limited in renal cancer imaging research \cite{ursprung2020radiomics}. For renal cancer, it therefore remains unclear whether the addition of a structured radiomics modality improves only the ranking performance of deep models or also the quality of the probabilities they produce.

In this study, we investigate this question through a controlled comparison of image-only and radiomics--image fusion models for binary ccRCC classification across various combinations of neural network architectures and radiomics fusion methods. 

\section{Methods}


To systematically evaluate the effect of radiomics--image fusion, we used a structured $4\times5$ model evaluation grid comprising four image model families and five prediction configurations within each family: an image-only baseline, three radiomics-fusion variants, and a probability-average ensemble of the three fusion models. The specific image backbones and fusion mechanisms are described in Section~\ref{sec:models}. A radiomics-only SVC was evaluated separately as an additional baseline.

\subsection{Dataset and Preprocessing}

The study used the Kidney Tumour Segmentation 2023 (KiTS23) dataset
\cite{heller2023kits}, a multi-institutional collection of contrast-enhanced abdominal CT scans with kidney and renal-tumour annotations. The classification task was formulated as clear-cell renal cell carcinoma (ccRCC) versus non-clear-cell RCC. Multifocal cases were excluded to avoid ambiguity between patient- and lesion-level labels. A total of 396 patients were included. A fixed patient-level split was used consistently across all model families, with approximately 60\% of cases assigned to training, 25\% to validation, and 15\% to held-out testing. The validation and test partitions had the same class distributions, with ccRCC accounting for approximately 70\% of cases in both partitions. The validation partition was used for model selection, whereas the held-out test partition was used only for final evaluation and was not used for model fitting or checkpoint selection.

All CT volumes and tumour masks were reoriented to RAS orientation and resampled to 1.0-mm isotropic spacing. CT intensities were clipped to $[-150,200]$ HU. Linear interpolation was used for images and nearest-neighbour interpolation for segmentation masks \cite{zwanenburg2020image}. A zero-margin tumour region of interest was used for radiomics extraction, whereas image-based models used a tumour-centred context region with a 12-voxel margin. When multiple tumour annotations were available, a STAPLE consensus mask was generated \cite{warfield2004simultaneous}. The underlying patients, labels, and split membership were kept identical across model comparisons.

\subsection{Radiomics Representation}

Radiomics features were extracted from the tumour region using PyRadiomics \cite{van2017computational} with IBSI-aligned definitions \cite{zwanenburg2020image}. Shape features were calculated from the original image, while first-order and texture features were extracted from both the original and Laplacian-of-Gaussian-filtered images with $\sigma\in\{1,2,3\}$ using a fixed bin width of 25 HU.

The initial 386 features were reduced by removing features with ICC $<0.75$, the lowest 20\% by median absolute deviation, and highly correlated features with $|\rho|>0.95$, followed by hierarchical clustering with a distance threshold of 0.2. This resulted in a 73-dimensional radiomics representation. For fusion models, radiomics features were standardised using statistics estimated from the training partition and applied unchanged to the validation and test partitions. A support vector classifier (SVC) using the selected radiomics features was included as a radiomics-only baseline.

\subsection{Image Representations and Classification Models}
\label{sec:models}

We evaluated conventional convolutional and pretrained image representations alongside MedVAE representations~\cite{varma2026medvae}. The conventional image models used a 3D ResNet-18 architecture~\cite{he2016deep} either trained from random initialisation or initialised from MedicalNet weights \cite{chen2019med3d}. For each backbone, an image-only classifier and three corresponding radiomics-fusion variants, concatenation~\cite{huang2020fusion}, cross-attention~\cite{tsai2019multimodal}, and gated fusion~\cite{arevalo2017gated}, were evaluated under the same patient-level split and training protocol. For fusion models, the image and radiomics representations were projected into compatible latent spaces before fusion. 

MedVAE~\cite{varma2026medvae} 
was evaluated in both 2D and 3D configurations. For the 2D models, tumour-containing axial slices were selected according to tumour area and the top 11 slices were encoded individually. Their latent representations were projected and mean-pooled to obtain a case-level image representation. For the 3D models, the tumour-centred CT volume was encoded directly using the volumetric MedVAE encoder, and the resulting latent representation was flattened and projected to a case-level embedding. Image-only 2D and 3D MedVAE classifiers were evaluated alongside their radiomics-fusion counterparts.

\paragraph{Concatenation.}
The projected image and radiomics embeddings were directly concatenated~\cite{huang2020fusion},
\begin{equation}
    \mathbf{h}_{\mathrm{concat}}
    =
    [\mathbf{f}_{\mathrm{img}};\mathbf{h}_{\mathrm{rad}}],
\end{equation}
and the joint representation was passed to a classification head producing two logits. This provides a simple fusion baseline in which the classifier learns the joint contribution of both modalities without an explicit cross-modal interaction mechanism.

\vspace{-0.1in} 

\paragraph{Cross-attention.}
Cross-attention was used to model conditional interactions between the image and radiomics representations~\cite{tsai2019multimodal}. Image token(s) were used as queries, whereas the projected radiomics representation was transformed into a small set of radiomics tokens serving as keys and values. For the 2D model, let $\mathbf{U}\in\mathbb{R}^{k\times d}$ denote the $k$ slice-level image embeddings and let $\mathbf{h}_{\mathrm{rad}}\in\mathbb{R}^{d}$ denote the projected radiomics representation. The image queries and radiomics tokens were constructed as
\begin{equation}
    \mathbf{Q}=\mathbf{U}\mathbf{W}_{Q},
    \qquad
    \mathbf{R}=
    \mathrm{reshape}
    \left(
    \mathbf{h}_{\mathrm{rad}}\mathbf{W}_{R}
    \right),
\end{equation}
where $\mathbf{Q}\in\mathbb{R}^{k\times d_c}$ and
$\mathbf{R}\in\mathbb{R}^{T\times d_c}$. Cross-modal attention was then computed as
\begin{equation}
    \mathrm{Attn}(\mathbf{Q},\mathbf{R},\mathbf{R})
    =
    \mathrm{softmax}
    \left(
    \frac{\mathbf{Q}\mathbf{R}^{\top}}{\sqrt{d_c}}
    \right)\mathbf{R}.
    \label{eq:cross_attention}
\end{equation}
The attended image tokens were further processed using residual connections, layer normalisation, and a feed-forward block, then mean-pooled and combined with the pooled image representation for classification. For ResNet-18, MedicalNet, and 3D MedVAE, the projected case-level image representation served as a single query under the same formulation. Thus, the radiomics tokens condition the image representation before the final classification stage rather than being combined only by late feature concatenation.

\paragraph{Gated fusion.}
For gated fusion, the projected image representation
$\mathbf{f}_{\mathrm{img}}\in\mathbb{R}^{d}$ and radiomics representation $\mathbf{h}_{\mathrm{rad}}\in\mathbb{R}^{d}$ were combined using a learned feature-wise gate~\cite{arevalo2017gated}:
\begin{equation}
    \mathbf{g}
    =
    \sigma\left(\mathrm{MLP}(\mathbf{h}_{\mathrm{rad}})\right),
\end{equation}
\begin{equation}
    \mathbf{h}_{\mathrm{fuse}}
    =
    \mathbf{g}\odot\mathbf{f}_{\mathrm{img}}
    +
    (1-\mathbf{g})\odot\mathbf{h}_{\mathrm{rad}},
    \label{eq:gated_fusion}
\end{equation}
where $\odot$ denotes element-wise multiplication. The fused representation was subsequently passed to a linear classifier producing two class logits. For the 3D MedVAE gated model, both modalities were projected into a 512-dimensional shared space and the gate was generated by a $512\rightarrow256\rightarrow512$ multilayer perceptron with ReLU and sigmoid activations.

\paragraph{Probability-average ensembles.} As a secondary robustness analysis, equal-weight probability-average ensembles were constructed separately for the ResNet-18, MedicalNet, 2D MedVAE, and 3D MedVAE families using the model-native predicted probabilities of ccRCC from concatenation, cross-attention, and gated fusion:

%
\begin{equation}
    p_{i,\mathrm{ensemble}}
    =
    \frac{
        p_{i,\mathrm{concat}}
        +
        p_{i,\mathrm{cross}}
        +
        p_{i,\mathrm{gated}}
    }{3},
    \qquad i=1,\ldots,n,
\end{equation}
\noindent where
$p_{i,m}=P_m(Y_i=1\mid \mathbf{x}_i)$ denotes the model-native
predicted probability of ccRCC for the $i$-th sample produced by
model $m \in \{\mathrm{concat},\mathrm{cross},\mathrm{gated}\}$,
and $p_{i,\mathrm{ensemble}}$ denotes the corresponding ensemble
probability. The ensemble weights were fixed a priori at $1/3$;
no weights were estimated from the data, and test set outcomes
were not used in ensemble construction.

\paragraph{Training Protocol.} All neural network models used the same fixed patient-level split and were trained with class-weighted cross-entropy and Adam optimisation. CT inputs were standardised over non-zero voxels, centre-cropped or padded to the architecture-specific size, and augmented using random flips, 90$^\circ$ rotations, and affine transformations. Validation AUC guided learning rate scheduling, early stopping, and checkpoint selection. ResNet-18 and MedicalNet used $128^3$ inputs, while 2D MedVAE used the 11 largest tumour-containing axial slices from a $128^3$ context volume and 3D MedVAE used $96^3$ inputs. MedVAE encoders were partially fine-tuned, with pretrained parameters assigned a lower learning rate than newly initialised downstream layers ($10^{-5}$ versus $10^{-4}$). All fusion models used the same 73-dimensional radiomics representation standardised using training-set statistics.
In total, 17 individual models and four probability-average ensembles were evaluated. Validation-selected checkpoints were frozen before final test set evaluation.

\subsection{Probability Calibration}

Calibration was evaluated using the model-native probability of ccRCC, $p_i=\widehat{P}(Y_i=1\mid x_i)$, where $Y_i=1$ denotes ccRCC. For neural networks, $p_i$ was obtained from the ccRCC component of the two-class softmax output. For the radiomics SVC, rather than softmax outputs, $p_i$ was obtained from the classifier's fitted class probability estimates, which use an internal probability mapping procedure.

\begin{sloppypar}
Probability calibration was quantified using expected calibration error (ECE)~\cite{guo2017calibration,silva2023classifier}, calibration intercept and slope~\cite{stevens2020validation}, and reliability diagrams, while the Brier score and NLL were used as complementary proper scoring rules for overall probabilistic performance. Proper scoring rules are instance-wise measures for evaluating the quality of probability estimates that avoid the need for score binning~\cite{silva2023classifier}.

To limit sparse bins in the relatively small test cohort, ECE was calculated using an equal-frequency binning scheme~\cite{silva2023classifier} as a scalar summary of calibration error:
\end{sloppypar}

\begin{equation}
\mathrm{ECE}
=
\sum_{m=1}^{M}
\frac{|B_m|}{n}
\left|
\overline{p}_m-\overline{y}_m
\right|,
\label{eq:ece}
\end{equation}
where $\overline{p}_m$ denotes the mean predicted probability and $\overline{y}_m$ the observed ccRCC proportion in bin $B_m$, $M$ denotes the number of the bins, and $n$ is the total number of samples. 
ECE was calculated using five equal-frequency bins ($M=5$). 

The Brier score and NLL are defined as
\begin{equation}
\mathrm{Brier}
=
\frac{1}{n}
\sum_{i=1}^{n}
(p_i-y_i)^2,
\label{eq:brier}
\end{equation}
and
\begin{equation}
\mathrm{NLL}
=
-\frac{1}{n}
\sum_{i=1}^{n}
\left[
y_i\log(p_i)
+
(1-y_i)\log(1-p_i)
\right],
\label{eq:nll}
\end{equation}
\noindent where $p_i$ and $y_i$ denote the predicted probability and the true label (according to the ground truth) of the $i$-th sample, respectively. 
Lower values for Brier score and NLL indicate better probabilistic prediction performance.
ECE was interpreted together with the Brier score and NLL rather than as a standalone measure \cite{glenn1950verification,van2019calibration}.

Calibration intercept and slope were estimated using logistic calibration regression \cite{van2019calibration,lambin2017radiomics,hatt2023joint,christiansen2025international}:
\begin{equation}
    \operatorname{logit}\{P(Y_i=1)\}
    =
    \alpha
    +
    \beta\operatorname{logit}(p_i),
\end{equation}
where $\alpha$ is the calibration intercept and $\beta$ is the calibration slope. Perfect calibration corresponds to $\alpha=0$ and $\beta=1$. The reported intercept was the joint intercept from the free-intercept/free-slope calibration model rather than calibration-in-the-large.

Reliability diagrams~\cite{guo2017calibration,murphy1977reliability} were used to plot the observed ccRCC proportion against the mean predicted probability in each of the five equal-frequency bins. Wilson 95\% confidence intervals were calculated for the observed event proportion. Points above the identity line indicate underconfidence, whereas points below the line indicate overconfidence.

The primary analysis comprised 12 matched image-only versus radiomics-fusion comparisons, with each fusion model compared against the corresponding image-only model using the same backbone, patient split, and evaluation protocol. Ninety-five percent confidence intervals were estimated using 1,000 ordinary patient-level bootstrap samples with percentile limits. The four probability-average ensembles were evaluated separately against their corresponding image-only models, and ensemble-minus-image-only differences in ECE, Brier score, and NLL were assessed using 1,000 paired patient-level bootstrap samples with identical resampled patient indices.




\section{Results and Discussion}

\subsection{Discrimination, Calibration, and Probabilistic Performance}
\begin{sloppypar}
Table~\ref{tab:combined_performance_calibration}
summarises performance across the 17 individual models and four probability-average ensembles. Across the 12 matched image-only versus radiomics-fusion comparisons, AUC and AP were higher for every fusion model. The Brier score and NLL were lower in 11 of 12 comparisons each, whereas ECE was lower in eight of 12. Thus, the discrimination gain associated with radiomics fusion was highly consistent across architectures, while improvements in the proper scoring rules were also frequent and reductions in calibration error were more fusion-dependent.
\end{sloppypar}

\begin{sloppypar}
Architecture-specific results showed that the fusion strategy with the strongest discrimination was not necessarily the one with the lowest calibration error. For ResNet-18, cross-attention achieved the highest discrimination, whereas concatenation yielded the lowest ECE and gated fusion the lowest Brier score and NLL. A similar divergence was observed for MedicalNet and 2D MedVAE. In contrast, 3D MedVAE gated fusion showed the strongest overall individual model profile, increasing AUC from 0.632 to 0.827 and AP from 0.772 to 0.922 while reducing ECE from 0.178 to 0.106, Brier score from 0.230 to 0.150, and NLL from 1.359 to 0.472. The radiomics-only SVC nevertheless achieved the lowest ECE (0.063).
\end{sloppypar}

Calibration regression provided a complementary view of probability calibration (Table~\ref{tab:combined_performance_calibration}). Radiomics fusion yielded slope point estimates closer to the ideal value of 1 for both CNN backbones and for 3D MedVAE, although confidence intervals were wide and the magnitude of improvement varied by fusion mechanism. The 2D MedVAE image-only model already had a slope near 1, and fusion produced mixed changes. These findings further indicate that calibration behaviour varied across image representations and fusion strategies.

\begin{table}[t]
\caption{Internal discrimination, probability calibration, probabilistic performance, and calibration-regression parameters on the held-out KiTS23 test cohort ($n=60$). Values in brackets denote 95\% ordinary patient-bootstrap confidence intervals. Ideal calibration corresponds to slope $=1$ and intercept $=0$.}
\label{tab:combined_performance_calibration}
\centering
\scriptsize
\renewcommand{\arraystretch}{1.08} 
\begin{tabular}{llccccccc}
\hline
\shortstack{Image\\Model} & Radiomics & AUC & AP & ECE & Brier & NLL & \shortstack{Calibration\\slope} & \shortstack{Calibration\\intercept} \\
\hline
- & radiomics only & 0.766 & 0.894 & 0.063 & 0.186 & 0.550 & 0.91 [0.26, 2.25] & -0.02 [-1.47, 0.98] \\
\hline
\multirow{5}{*}{\rotatebox{90}{\tiny ResNet18}} & image only & 0.532 & 0.738 & 0.153 & 0.216 & 0.646 & 0.20 [-0.59, 1.21] & 0.62 [-0.35, 1.73] \\
 & concatenation & 0.663 & 0.799 & \textbf{0.096} & 0.207 & 0.607 & 0.61 [-0.03, 1.52] & 0.56 [-0.04, 1.28] \\
 & cross-attention & \textbf{0.790} & \textbf{0.897} & 0.136 & 0.190 & 0.556 & 0.79 [0.38, 1.51] & 0.78 [0.15, 1.65] \\
 & gated fusion & 0.784 & 0.865 & 0.124 & \textbf{0.176} & \textbf{0.533} & 1.22 [0.53, 2.32] & 0.61 [0.02, 1.36] \\
 & ensemble & 0.780 & 0.879 & 0.119 & 0.178 & 0.537 & 1.18 [0.56, 2.31] & 0.56 [-0.11, 1.30] \\
\hline
\multirow{5}{*}{\rotatebox{90}{\tiny MedicalNet}} & image only & 0.639 & 0.852 & 0.206 & 0.259 & 0.741 & 0.20 [-0.56, 0.59] & 0.86 [0.34, 1.53] \\
 & concatenation & 0.746 & 0.866 & 0.129 & 0.188 & 0.559 & 1.70 [0.75, 3.51] & 0.41 [-0.20, 1.11] \\
 & cross-attention & 0.776 & 0.902 & 0.219 & 0.221 & 0.652 & 0.80 [0.41, 1.60] & 1.15 [0.59, 2.01] \\
 & gated fusion & 0.786 & 0.864 & \textbf{0.115} & \textbf{0.169} & \textbf{0.518} & 1.14 [0.51, 2.33] & 0.57 [-0.07, 1.35] \\
 & ensemble & \textbf{0.815} & \textbf{0.905} & 0.156 & 0.178 & 0.537 & 1.64 [0.90, 3.53] & 0.77 [0.12, 1.62] \\
\hline
\multirow{5}{*}{\rotatebox{90}{\tiny 2D MedVAE}} & image only & 0.595 & 0.788 & 0.134 & 0.221 & 0.650 & 1.11 [-0.47, 2.81] & -0.97 [-3.78, 1.62] \\
 & concatenation & 0.747 & 0.880 & 0.149 & 0.220 & 0.669 & 0.42 [0.17, 0.89] & 0.72 [0.14, 1.48] \\
 & cross-attention & 0.761 & 0.875 & \textbf{0.065} & 0.171 & 0.532 & 0.62 [0.34, 1.14] & 0.06 [-0.70, 0.79] \\
 & gated fusion & \textbf{0.796} & \textbf{0.901} & 0.174 & 0.200 & 0.604 & 1.02 [0.56, 1.94] & -1.35 [-3.13, -0.15] \\
 & ensemble & 0.792 & 0.899 & 0.071 & \textbf{0.162} & \textbf{0.492} & 0.94 [0.48, 1.80] & -0.16 [-0.99, 0.63] \\
\hline
\multirow{5}{*}{\rotatebox{90}{\tiny 3D MedVAE}} & image only & 0.632 & 0.772 & 0.178 & 0.230 & 1.359 & 0.07 [-0.05, 0.35] & 0.50 [-0.39, 1.35] \\
 & concatenation & 0.738 & 0.863 & 0.121 & 0.194 & 0.624 & 0.41 [0.17, 0.80] & 0.31 [-0.47, 1.02] \\
 & cross-attention & 0.795 & 0.889 & 0.201 & 0.241 & 1.204 & 0.16 [0.08, 0.30] & 0.37 [-0.31, 1.02] \\
 & gated fusion & \textbf{0.827} & \textbf{0.922} & \textbf{0.106} & \textbf{0.150} & \textbf{0.472} & 0.52 [0.34, 0.93] & 0.02 [-0.89, 0.72] \\
 & ensemble & 0.807 & 0.911 & 0.140 & 0.165 & 0.494 & 0.59 [0.36, 0.96] & 0.14 [-0.59, 0.87] \\
\hline
\end{tabular}
\end{table}

\subsection{Calibration and Probabilistic Performance Across Fusion Strategies}

The expanded comparison across four image backbones revealed consistent but non-uniform effects of the three fusion strategies. Gated fusion reduced the Brier score and NLL relative to the corresponding image-only model in all four backbones and reduced ECE in three of four. Concatenation reduced the Brier score in all four backbones and reduced both ECE and NLL in three of four. Cross-attention reduced NLL in all four backbones and Brier score in three of four, but reduced ECE in only two of four. Thus, improvements in the proper scoring rules were more reproducible across fusion mechanisms than reductions in calibration error.

The fusion strategy that maximised discrimination was not necessarily the one with the lowest calibration error. For ResNet-18, cross-attention achieved the highest AUC and AP, whereas concatenation produced the lowest ECE and gated fusion the lowest Brier score and NLL. MedicalNet cross-attention similarly improved discrimination while slightly increasing ECE relative to the image-only model. In contrast, 3D MedVAE gated fusion provided the most favourable joint profile across discrimination, calibration error, and proper scoring rules. These findings indicate that the probabilistic consequences of radiomics fusion cannot be inferred from discrimination alone and depend on how the two modalities are integrated. This variation across individual fusion mechanisms motivated the ensemble analysis, which tested whether averaging their complementary predictions could yield a more consistent performance profile.


\subsection{Probability-Average Ensemble Robustness Analysis}

To examine whether the broader probabilistic-performance pattern depended on selecting a single fusion mechanism, equal-weight probability-average ensembles of concatenation, cross-attention, and gated fusion were compared with the corresponding image-only models. Importantly, this simple averaging strategy improved all five primary performance metrics at the point-estimate level across all four model families: AUC and AP increased, while ECE, Brier score, and NLL decreased (Table \ref{tab:combined_performance_calibration}
).

Paired patient-bootstrap differences for the probability-average ensembles are summarised in Table~\ref{tab:ensemble_differences}. 
Negative differences indicate lower probabilistic error for the ensemble.
None of the ECE confidence intervals excluded zero, although the point estimates favoured the ensemble across all four model families. Notably, for 3D MedVAE, probability averaging reduced ECE from 0.178 to 0.140, Brier score from 0.230 to 0.165, and NLL from 1.359 to 0.494, corresponding to relative reductions of 21.3\%, 28.3\%, and 63.6\%, respectively. Thus, averaging the three complementary fusion strategies produced consistently favourable point estimates across discrimination, calibration error, and overall probabilistic performance, while paired-bootstrap evidence was stronger for the proper scoring rules than for ECE.

\begin{table}[t]
\caption{Paired patient-bootstrap differences between probability-average ensembles and their corresponding image-only models. Differences are defined as ensemble minus image-only; negative values therefore indicate lower probabilistic error for the ensemble. Values in brackets denote 95\% paired-bootstrap confidence intervals. Bold values indicate intervals excluding zero.}
\label{tab:ensemble_differences}
\centering
\scriptsize
\setlength{\tabcolsep}{3.5pt}
\renewcommand{\arraystretch}{1.05}
\begin{tabular}{lccc}
\hline
Model family &
$\Delta$ECE (95\% CI) &
$\Delta$Brier (95\% CI) &
$\Delta$NLL (95\% CI) \\
\hline
ResNet18 &
$-0.034$ [$-0.122$, 0.124] &
$-0.037$ [$-0.105$, 0.028] &
$-0.109$ [$-0.286$, 0.060] \\

MedicalNet &
$-0.050$ [$-0.175$, 0.040] &
$\textbf{-0.080}$ [$\textbf{-0.136}$, $\textbf{-0.022}$] &
$\textbf{-0.204}$ [$\textbf{-0.354}$, $\textbf{-0.061}$] \\

2D MedVAE &
$-0.063$ [$-0.133$, 0.043] &
$\textbf{-0.059}$ [$\textbf{-0.115}$, $\textbf{-0.008}$] &
$\textbf{-0.157}$ [$\textbf{-0.293}$, $\textbf{-0.031}$] \\

3D MedVAE &
$-0.038$ [$-0.163$, 0.070] &
$-0.066$ [$-0.151$, 0.012] &
$\textbf{-0.864}$ [$\textbf{-1.783}$, $\textbf{-0.163}$] \\
\hline
\end{tabular}
\end{table}





\begin{figure}[!t]
    \centering
    \includegraphics[width=0.82\textwidth,trim=10 8 10 8,clip]{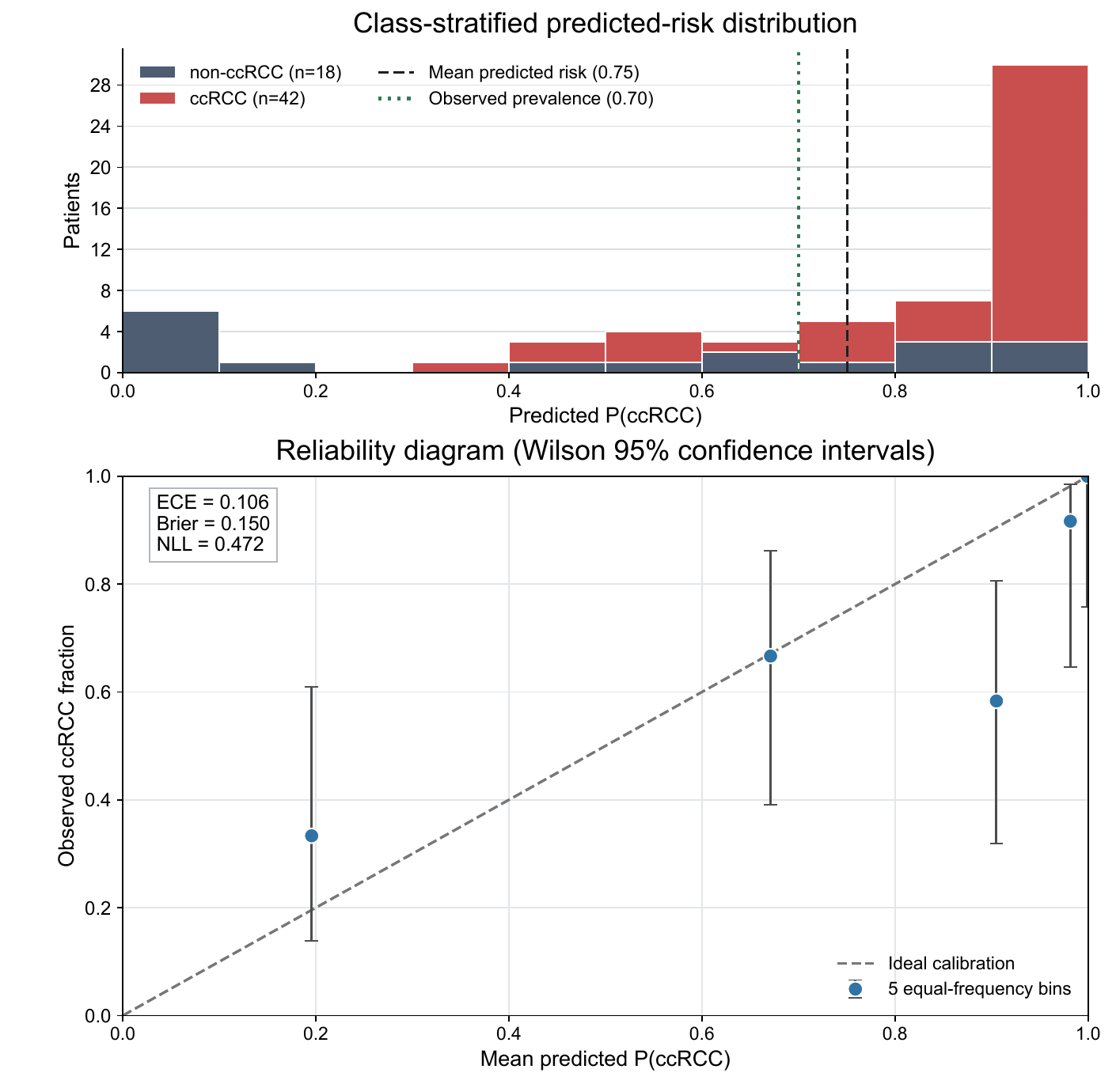}
    \vspace{-0.1in} 
    \caption{Predicted-risk distribution and reliability diagram for the 3D MedVAE gated-fusion model on the held-out KiTS23 test cohort. Reliability was evaluated using five equal-frequency bins; error bars denote Wilson 95\% confidence intervals and the dashed line represents perfect calibration.}
    \label{fig:primary_reliability}
\end{figure}

\subsection{Reliability of the 3D MedVAE Gated-Fusion Model}

The predicted-risk distribution and reliability diagram for 3D MedVAE gated fusion are shown in Figure~\ref{fig:primary_reliability}. Predictions were concentrated toward high ccRCC probabilities and showed heterogeneous deviations from the identity line across probability bins. The model achieved ECE 0.106, Brier score 0.150, and NLL 0.472. Calibration regression yielded an intercept of 0.021 [95\% CI: $-0.890$, 0.722] and a slope of 0.517 [0.340, 0.930]. The near-zero joint intercept suggested limited overall probability offset after accounting for the slope, whereas the slope below 1 indicated that the predicted probabilities remained overly extreme across the probability range.



\FloatBarrier
\section{Conclusion}
\vspace{-0.06cm} 
Radiomics--image fusion improved discrimination consistently across four image-model families, while improvements in overall probabilistic performance were more reproducible than reductions in calibration error. Across 12 matched comparisons, AUC and AP were higher in all 12 comparisons, Brier score and NLL were lower in 11 of 12, and ECE was lower in eight of 12. Calibration behaviour nevertheless varied across image architectures and fusion mechanisms. Among individual models, 3D MedVAE gated fusion achieved the strongest joint discrimination--probabilistic performance profile. Probability averaging across the three fusion strategies also produced favourable point estimates relative to image-only models across all four families, with stronger paired-bootstrap support for Brier score and NLL than for ECE. These results support evaluating multimodal medical-image classifiers using complementary discrimination, calibration, and proper-scoring measures rather than discrimination alone. External multi-centre validation is required to determine the generalisability of these findings.

\bibliographystyle{splncs04}
\bibliography{bib}

@article{bex2025european,
  title   = {European Association of Urology Guidelines on Renal Cell Carcinoma: The 2025 Update},
  author  = {Bex, Axel and {Abu Ghanem}, Yasmin and Albiges, Laurence and Bonn, Stephanie and Campi, Riccardo and Capitanio, Umberto and Dabestani, Saeed and Hora, Milan and Klatte, Tobias and Kuusk, Teele and others},
  journal = {European Urology},
  volume  = {87},
  number  = {6},
  pages   = {683--696},
  year    = {2025},
  doi     = {10.1016/j.eururo.2025.02.020}
}

@article{wang2021radiomics,
  title   = {Radiomics Models Based on Enhanced Computed Tomography to Distinguish Clear Cell from Non-Clear Cell Renal Cell Carcinomas},
  author  = {Wang, Ping and Pei, Xu and Yin, Xiao-Ping and Ren, Jia-Liang and Wang, Yun and Ma, Lu-Yao and Du, Xiao-Guang and Gao, Bu-Lang},
  journal = {Scientific Reports},
  volume  = {11},
  number  = {1},
  pages   = {13729},
  year    = {2021},
  doi     = {10.1038/s41598-021-93069-z}
}

@article{zwanenburg2020image,
  title={The image biomarker standardization initiative: standardized quantitative radiomics for high-throughput image-based phenotyping},
  author={Zwanenburg, Alex and Valli{\`e}res, Martin and Abdalah, Mahmoud A and Aerts, Hugo JWL and Andrearczyk, Vincent and Apte, Aditya and Ashrafinia, Saeed and Bakas, Spyridon and Beukinga, Roelof J and Boellaard, Ronald and others},
  journal={Radiology},
  volume={295},
  number={2},
  pages={328--338},
  year={2020},
  publisher={Radiological Society of North America}
}

@article{uhm2021deep,
  title={Deep learning for end-to-end kidney cancer diagnosis on multi-phase abdominal computed tomography},
  author={Uhm, Kwang-Hyun and Jung, Seung-Won and Choi, Moon Hyung and Shin, Hong-Kyu and Yoo, Jae-Ik and Oh, Se Won and Kim, Jee Young and Kim, Hyun Gi and Lee, Young Joon and Youn, Seo Yeon and others},
  journal={NPJ precision oncology},
  volume={5},
  number={1},
  pages={54},
  year={2021},
  publisher={Nature Publishing Group UK London}
}

@article{van2019calibration,
  title   = {Calibration: The Achilles Heel of Predictive Analytics},
  author  = {Van Calster, Ben and McLernon, David J. and van Smeden, Maarten and Wynants, Laure and Steyerberg, Ewout W. and others},
  journal = {BMC Medicine},
  volume  = {17},
  number  = {1},
  pages   = {230},
  year    = {2019},
  doi     = {10.1186/s12916-019-1466-7}
}

@inproceedings{guo2017calibration,
  title     = {On Calibration of Modern Neural Networks},
  author    = {Guo, Chuan and Pleiss, Geoff and Sun, Yu and Weinberger, Kilian Q.},
  booktitle = {Proceedings of the 34th International Conference on Machine Learning},
  series    = {Proceedings of Machine Learning Research},
  volume    = {70},
  pages     = {1321--1330},
  year      = {2017},
  publisher = {PMLR}
}

@article{murphy1977reliability,
  title={Reliability of subjective probability forecasts of precipitation and temperature},
  author={Murphy, Allan H and Winkler, Robert L},
  journal={Journal of the Royal Statistical Society Series C: Applied Statistics},
  volume={26},
  number={1},
  pages={41--47},
  year={1977},
  publisher={Oxford University Press}
}

@article{silva2023classifier,
  title={Classifier calibration: a survey on how to assess and improve predicted class probabilities},
  author={Silva Filho, Telmo and Song, Hao and Perello-Nieto, Miquel and Santos-Rodriguez, Raul and Kull, Meelis and Flach, Peter},
  journal={Machine Learning},
  volume={112},
  number={9},
  pages={3211--3260},
  year={2023},
  publisher={Springer}
}

@article{stevens2020validation,
  title={Validation of clinical prediction models: what does the “calibration slope” really measure?},
  author={Stevens, Richard J and Poppe, Katrina K},
  journal={Journal of clinical epidemiology},
  volume={118},
  pages={93--99},
  year={2020},
  publisher={Elsevier}
}

@article{nelson2026gaussian,
  title={Gaussian Meta-Space Augmentation for Stacking Ensembles in Multimodal IPMN Risk Stratification},
  author={Nelson, Max A and Tasci, Eminenur Sen and Wang, Zhixiang and Zhou, Zongwei and Aktas, Halil Ertugrul and Bejar, Andrea M and Keles, Elif and Hong, Ziliang and Taflan, S{\i}tk{\i} Safa and Tasci, Muhammed Enes and others},
  journal={arXiv preprint arXiv:2608.11472},
  year={2026}
}

@article{xi2020deep,
  title   = {Deep Learning to Distinguish Benign from Malignant Renal Lesions Based on Routine {MR} Imaging},
  author  = {Xi, Ianto Lin and Zhao, Yijun and Wang, Robin and Chang, Marcello and Purkayastha, Subhanik and Chang, Ken and Huang, Raymond Y. and Silva, Alvin C. and Vallieres, Martin and Habibollahi, Peiman and others},
  journal = {Clinical Cancer Research},
  volume  = {26},
  number  = {8},
  pages   = {1944--1952},
  year    = {2020}
}

@misc{raposo2026imaging,
  author  = {Raposo, Hugo},
  title   = {Is Imaging Redundant? Leakage-Controlled Multi-Representation Assessment of CT-Clinical Fusion for Two-Year Survival Prediction in Non-Small-Cell Lung Cancer},
  howpublished = {(June 29, 2026). Available at SSRN: \url{https://ssrn.com/abstract=7023198}}
}

@misc{
hu2026robust,
title={Robust Liver Fibrosis Staging under Scanner Shift with Deep and Radiomic Streams},
author={Shu Hu and Xin Wang},
year={2026},
url={https://openreview.net/forum?id=i3ainP2BVH}
}

@misc{heller2023kits,
  title  = {2023 Kidney and Kidney Tumor Segmentation Challenge},
  author = {Heller, Nicholas and Isensee, Fabian and Tejpaul, Rohit and Wood, Austin and Papanikolopoulos, Nikolaos and Weight, Christopher},
  year   = {2023},
  publisher = {Zenodo},
  doi    = {10.5281/zenodo.7840134}
}

@article{van2017computational,
  title   = {Computational Radiomics System to Decode the Radiographic Phenotype},
  author  = {Van Griethuysen, Joost J. M. and Fedorov, Andriy and Parmar, Chintan and Hosny, Ahmed and Aucoin, Nicole and Narayan, Vivek and Beets-Tan, Regina G. H. and Fillion-Robin, Jean-Christophe and Pieper, Steve and Aerts, Hugo J. W. L.},
  journal = {Cancer Research},
  volume  = {77},
  number  = {21},
  pages   = {e104--e107},
  year    = {2017},
  doi     = {10.1158/0008-5472.CAN-17-0339}
}

@article{warfield2004simultaneous,
  title={Simultaneous truth and performance level estimation (STAPLE): an algorithm for the validation of image segmentation},
  author={Warfield, Simon K and Zou, Kelly H and Wells, William M},
  journal={IEEE transactions on medical imaging},
  volume={23},
  number={7},
  pages={903--921},
  year={2004},
  publisher={IEEE}
}

@article{chen2019med3d,
  title={{Med3D}: Transfer learning for 3d medical image analysis},
  author={Chen, Sihong and Ma, Kai and Zheng, Yefeng},
  journal={arXiv preprint arXiv:1904.00625},
  year={2019}
}

@article{glenn1950verification,
  title={Verification of Forecasts Expressed in Terms of Probability},
  author={Brier, Glenn W.},
  journal={Monthly Weather Review},
  volume={78},
  number={1},
  pages={1--3},
  year={1950}
}

@article{ursprung2020radiomics,
  title   = {Radiomics of Computed Tomography and Magnetic Resonance Imaging in Renal Cell Carcinoma---A Systematic Review and Meta-Analysis},
  author  = {Ursprung, Stephan and Beer, Lucian and Bruining, Annemarie and Woitek, Ramona and Stewart, Grant D. and Gallagher, Ferdia A. and Sala, Evis},
  journal = {European Radiology},
  volume  = {30},
  number  = {6},
  pages   = {3558--3566},
  year    = {2020},
  doi     = {10.1007/s00330-020-06666-3}
}

@article{bhandari2021ct,
  title   = {{CT}-Based Radiomics for Differentiating Renal Tumours: A Systematic Review},
  author  = {Bhandari, Abhishta and Ibrahim, Muhammad and Sharma, Chinmay and Liong, Rebecca and Gustafson, Sonja and Prior, Marita},
  journal = {Abdominal Radiology},
  volume  = {46},
  number  = {5},
  pages   = {2052--2063},
  year    = {2021}
}

@article{huang2020fusion,
  author  = {Huang, Shih-Cheng and Pareek, Anuj and Seyyedi, Sajjad and Banerjee, Imon and Lungren, Matthew P.},
  title   = {Fusion of Medical Imaging and Electronic Health Records Using Deep Learning: A Systematic Review and Implementation Guidelines},
  journal = {npj Digital Medicine},
  volume  = {3},
  number  = {1},
  pages   = {136},
  year    = {2020},
  doi     = {10.1038/s41746-020-00341-z}
}

@InProceedings{varma2026medvae,
  title     = {{MedVAE}: Efficient Automated Interpretation of Medical Images with Large-Scale Generalizable Autoencoders},
  author    = {Varma, Maya and Kumar, Ashwin and van der Sluijs, Rogier and Ostmeier, Sophie and Blankemeier, Louis and Chambon, Pierre Joseph Marcel and Bluethgen, Christian and Prince, Jip and Langlotz, Curtis and Chaudhari, Akshay S},
  booktitle = {Proceedings of the 8th International Conference on Medical Imaging with Deep Learning},
  publisher = {PMLR},
  volume    = {301},
  pages     = {1597--1626},
  year      = {2026}
}

@inproceedings{he2016deep,
  title={Deep residual learning for image recognition},
  author={He, Kaiming and Zhang, Xiangyu and Ren, Shaoqing and Sun, Jian},
  booktitle={Proceedings of the IEEE conference on computer vision and pattern recognition},
  pages={770--778},
  year={2016}
}

@inproceedings{tsai2019multimodal,
  author    = {Tsai, Yao-Hung Hubert and Bai, Shaojie and Liang, Paul Pu and Kolter, J. Zico and Morency, Louis-Philippe and Salakhutdinov, Ruslan},
  title     = {Multimodal Transformer for Unaligned Multimodal Language Sequences},
  booktitle = {Proceedings of the 57th Annual Meeting of the Association for Computational Linguistics},
  pages     = {6558--6569},
  year      = {2019},
  doi       = {10.18653/v1/P19-1656}
}

@inproceedings{arevalo2017gated,
  author    = {Arevalo, John and Solorio, Thamar and Montes-y-G{\'o}mez, Manuel and Gonz{\'a}lez, Fabio A.},
  title     = {Gated Multimodal Units for Information Fusion},
  booktitle = {International Conference on Learning Representations Workshop},
  year      = {2017}
}

@article{lambin2017radiomics,
  title={Radiomics: the bridge between medical imaging and personalized medicine},
  author={Lambin, Philippe and Leijenaar, Ralph TH and Deist, Timo M and Peerlings, Jurgen and De Jong, Evelyn EC and Van Timmeren, Janita and Sanduleanu, Sebastian and Larue, Ruben THM and Even, Aniek JG and Jochems, Arthur and others},
  journal={Nature reviews Clinical oncology},
  volume={14},
  number={12},
  pages={749--762},
  year={2017},
  publisher={Nature Publishing Group UK London}
}

@article{hatt2023joint,
  title={Joint EANM/SNMMI guideline on radiomics in nuclear medicine: Jointly supported by the EANM Physics Committee and the SNMMI Physics, Instrumentation and Data Sciences Council},
  author={Hatt, Mathieu and Krizsan, AK and Rahmim, A and Bradshaw, TJ and Costa, P Fragoso and Forg{\'a}cs, Attila and Seifert, Robert and Zwanenburg, A and El Naqa, I and Kinahan, PE and others},
  journal={European Journal of Nuclear Medicine and Molecular Imaging},
  volume={50},
  number={2},
  pages={352--375},
  year={2023},
  publisher={Springer}
}

@article{christiansen2025international,
  title={International multicenter validation of AI-driven ultrasound detection of ovarian cancer},
  author={Christiansen, Filip and Konuk, Emir and Ganeshan, Adithya Raju and Welch, Robert and Pales Huix, Joana and Czekierdowski, Artur and Leone, Francesco Paolo Giuseppe and Haak, Lucia Anna and Fruscio, Robert and Gaurilcikas, Adrius and others},
  journal={Nature medicine},
  volume={31},
  number={1},
  pages={189--196},
  year={2025},
  publisher={Nature Publishing Group US New York}
}
\end{document}